\documentclass[11pt,a4]{article} 

\pdfoutput=1
\usepackage[utf8]{inputenc} 
\usepackage{graphicx}
\usepackage{wrapfig}
\usepackage[margin=10pt,font=small,labelfont=bf,
labelsep=endash]{caption}

\usepackage{geometry} 
\usepackage{graphicx} 

\usepackage{booktabs} 
\usepackage{array} 
\usepackage{paralist} 
\usepackage{verbatim} 
\usepackage{subfig} 

\usepackage{fancyhdr} 
\usepackage{sectsty}
\allsectionsfont{\sffamily\mdseries\upshape} 

\usepackage[nottoc,notlof,notlot]{tocbibind} 
\usepackage[titles,subfigure]{tocloft} 

\title{The Neutron and the Universe - History of a Relationship}

\author{Stephan Paul\\
        Physik Department E18, Technische Universit\"at M\"unchen\\
        James Franck Str.  D-85748 Garching \\
        E-mail: stephan.paul@tum.de } 
        
\begin{document}
\maketitle

\begin{abstract} We discuss selected topics in the field of particle- and astrophysics with neutrons. They have a direct link with our understanding of the history of the Universe and are related to recent, ongoing or future measurements. They deal with the structure of space-time (tests of gravitation at small distance scales), search for an electric dipole moment of the neutron (CP-violation and the origin of matter in the Universe), the neutron lifetime (rate of primordial nucleosynthesis) and the two-body decay of the neutron testing the V--A structure of weak interaction (right-handed neutrinos and the very early Universe). We describe the status, measurement methods and highlight experimental challenges.


\end{abstract}

\section{Introduction}
\begin{figure}\begin{centering}
         \includegraphics[width=13cm]{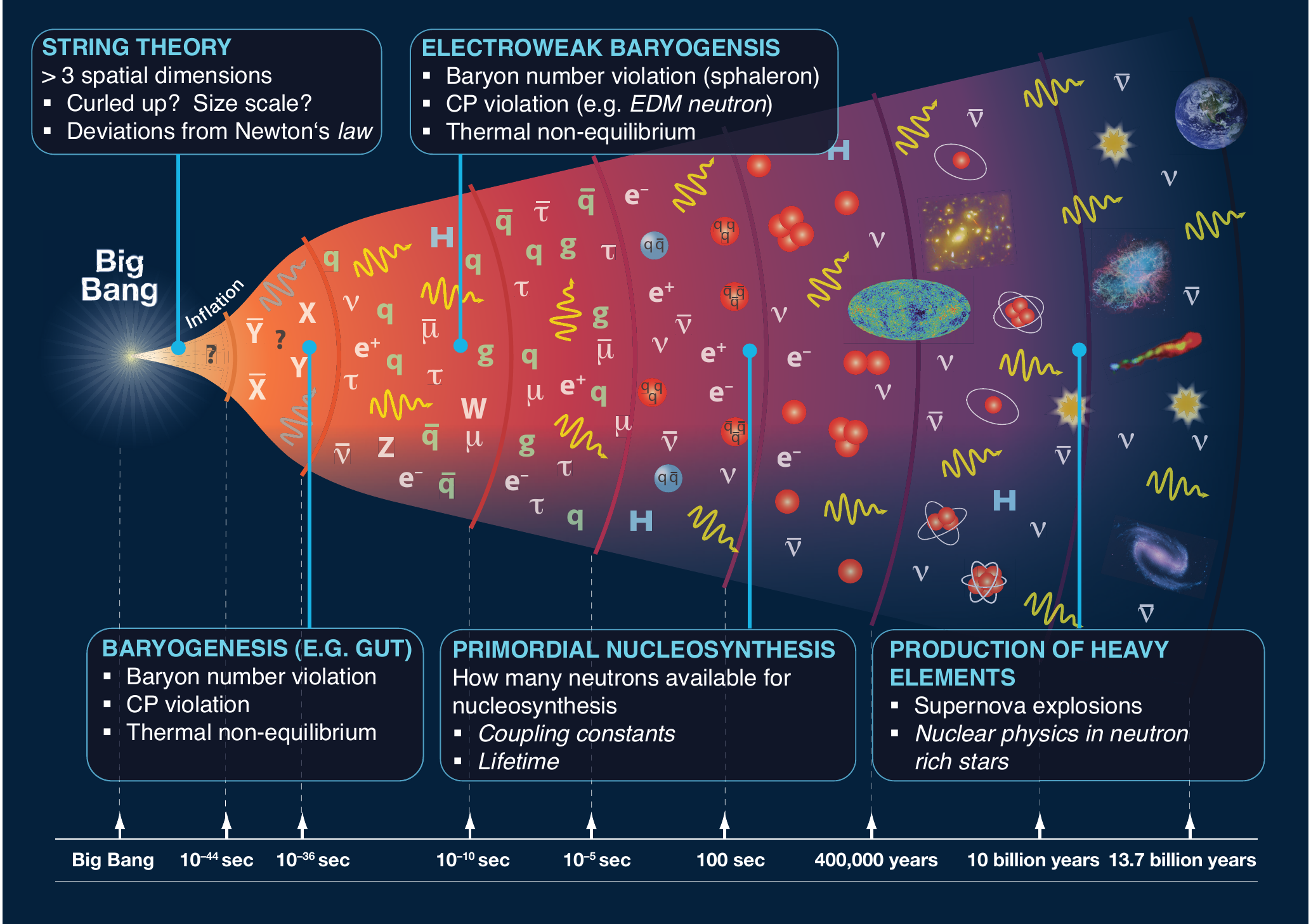}\\
         \caption{Illustration of the structure and development of the Universe in space (vertical axis) and time (horizontal axis). The time axis is quasi-logarithmic. The diagram depicts the different phases of the Universe corresponding to the standard model of cosmology. Many features are strongly model based and lack experimental support. Also indicated is the role of measurements performed with neutrons and their impact on supporting our present conception.}\label{Universe}\end{centering}
      \end{figure}

Looking at the time evolution of our Universe we can spot several epochs characterized by physical phases or phase transitions which left their imprint on the structure and functioning of our Universe, as we know it today (see fig.\ref{Universe}). While particle physics dominates for the first three minutes, gravitation has been dominant in shaping the Universe and creating the structures since. With the exception of nuclear physics processes being responsible for the creation of (heavy) elements and the life and death of stars, the later stage has no direct relation to particle physics.
\par
At times before 10$^{-15}$s our understanding of particle physics is based on extrapolation of existing models or new theoretical concepts which still lack experimental verification. The earliest times are the domain of string theory, which requires additional spatial dimensions, presumably curled up during the inflationary stage in the expansion of the Universe. No real guidance is given as to the length scale of those as of today but the effect should be noticeable by a deviation of the gravitational force from Newton's law at small distances or even additional short range gravitational like forces ($5^{th}$ force). Neutrons are an excellent tool for investigating this domain, presently exploring distance scales at order $\mu$\rm{m}.  \newline

Baryogenesis, reflected in the disappearance of anti-matter in the Universe, requires three ingredients: namely baryon number violation, CP-violation and thermal non-equilibrium \cite{sakharov}, each of which has no sound experimental verification. The inflationary epoch with its rapid spatial expansion of the Universe offers such strong thermal non-equilibrium and this concept is in agreement with cosmological observations and modeling of structures and the expansion of our Universe. It is thus the favoured time for baryogenesis.  If it is originating from subtle processes in the quark sector then baryon number violation and sizable CP-violation in these processes are necessary to accommodate for this large effect observed ($\approx 10^{-9}$). As flavour non-diagonal processes offer little room for new and large CP-violating effects and are (as for now) explained by the CKM-matrix, flavour diagonal effects are less constrained as could be revealed by a sizable electric dipole moment of nuclei or the neutron. Electroweak baryogenesis is another possibility (at times $\sim$ 10$^{-12}${\rm s}), but  less favoured by theorists owing to the presumably low mass of the Higgs-boson and thus the lack of a suitable phase transition). Alternative scenarios for the matter/antimatter asymmetry observed in the Universe are based on Leptogenesis scenarios not discussed here.\newline
Following the quark-gluon to hadron phase transition the weak interaction keeps a balance between protons and neutrons. Owing to the cooling of the Universe and the mass difference of the proton and the neutron, the equilibrium, which follows a Boltzmann distribution, is changed and reaches a value of about 6:1 in favour of protons. This value is governed by the weak nucleon coupling constants which are measured e.g. in neutron decay. Before strong fusion processes set in leading to the creation of deuterons and helium nuclei neutrons decay altering the ratio down to 7:1. This is the starting point for primordial nucleosynthesis processes (BBN). Thus neutron decay parameters (expressed in terms of the neutron lifetime) are input into BBN calculations and are a sensitive parameter within the standard model of cosmology.\\
As the Universe keeps expanding and cooling down the interaction rates become very small and the time evolution changes pace and becomes dominated by the gravitational force, thus leading to huge time scales. Only many billions of years later neutrons again appear as important players, namely in star burning processes with the production of heavier elements and in supernova explosions where neutron rich isotopes are key partners in the chain of r- and s-processes leading to the heaviest elements past iron. The understanding of these nuclei involve nuclear physics experiments with radioactive beams and shall not be discussed here. 

\noindent A detailed discussion of this topic can be found in \cite{Dubbers}.

\section{Tests of Gravity}
\begin{figure}\centering
         \includegraphics[width=8cm]{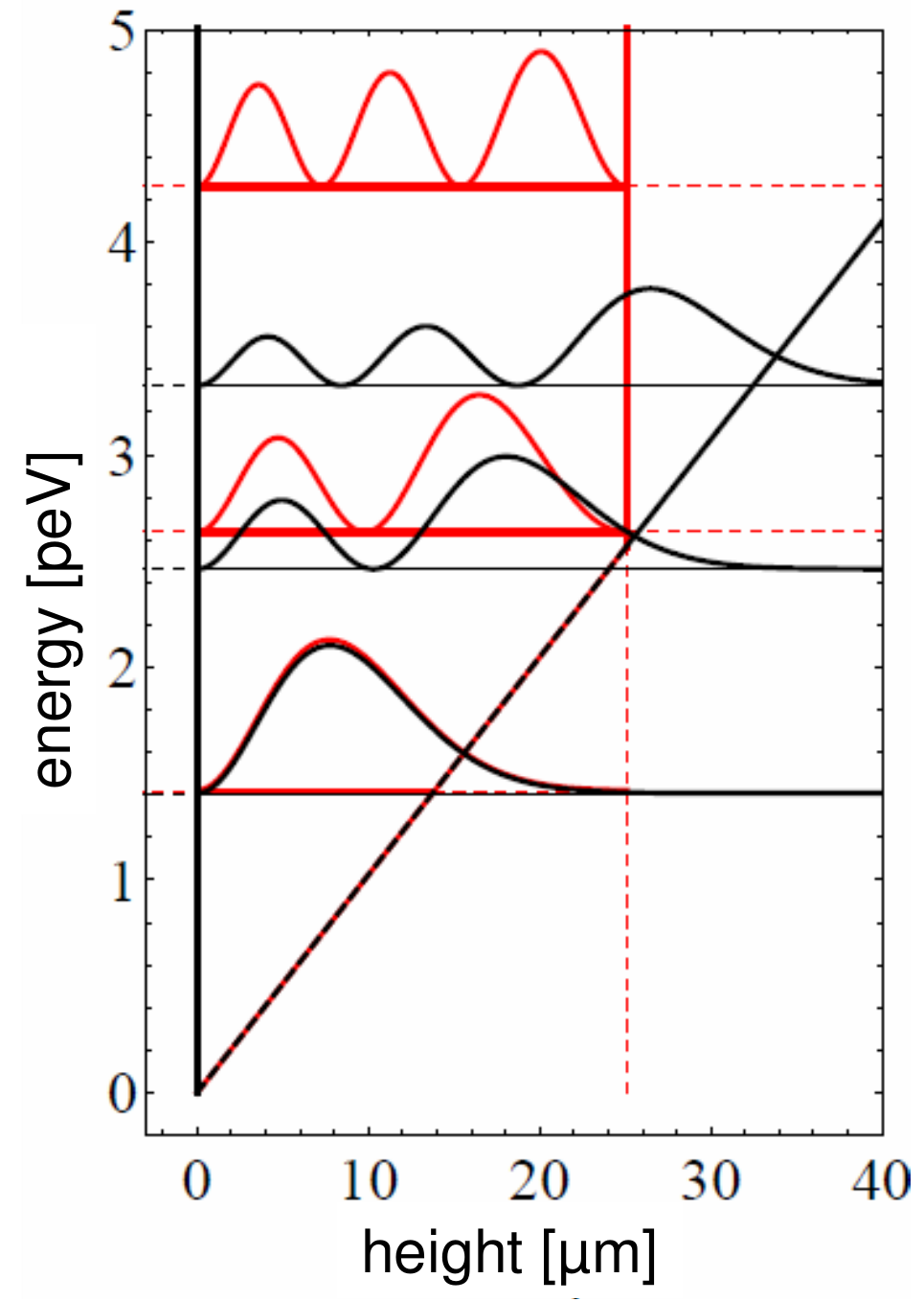}  \\       
         \caption{Solution of the Schr\"odinger equation for a linear potential $V_{{\rm grav}} = m\cdot g\cdot h_{{\rm abs}}$ (black lines). Modifications introduced by an absorber placed at height $h_{{\rm abs}}$ acting as boundary condition (red lines).}
         \label{airy}
      \end{figure}

Owing to their neutrality and their very small polarizability, neutrons are virtually immune to the effects of electric fields. Thus, in the absence of magnetic field gradients, they only feel gravitational forces (except at at distances of \AA\  from matter surfaces, where they are subject to the strong interaction in terms of effective Fermi-potentials). The interaction strength of the Earth's gravitational field with neutrons is about 100 neV/m. Thus, in order to probe gravitational effects neutrons of very small energy have to be used which exhibit a vertical kinetic energy component far below this value. \\
Neutrons with kinetic energies $\leq 250$ $\rm neV$ are called ultra cold neutrons (UCN). They are reflected at all angles of incidence from suitable mechanical surfaces, which exhibit effective repulsive Fermi potentials of 200-300 neV and thus act as perfect mirrors. \\
The vertical movement of very slow neutrons incident on such a mirror can be described by the Schr\"odinger equation with a linear potential ($V_{{\rm grav}} = m\cdot g\cdot h$) of which the solutions $\psi_n(h)$ are Airy functions. The movement is thus quantized and the probability to find a neutron above the mirror at a height {\it h} depends on the neutron energy which is related to the Eigenvalue {\it n} and the square $\left|{\psi_n(h)}\right|^2$ as depicted in fig.\ref{airy} (black lines). We can now place an absorber above the mirror at a height $h_{{\rm abs}}$ introducing boundary conditions which force the wave function to disappear at $h=h_{{\rm abs}}$, and thus modifying the solutions. This is depicted in fig.\ref{airy} (red lines).
This quantization in the vertical direction (no force is acting horizontally ) can now be probed experimentally and has been done in various steps.

\par
\begin{figure}\centering
         \includegraphics[width=7cm]{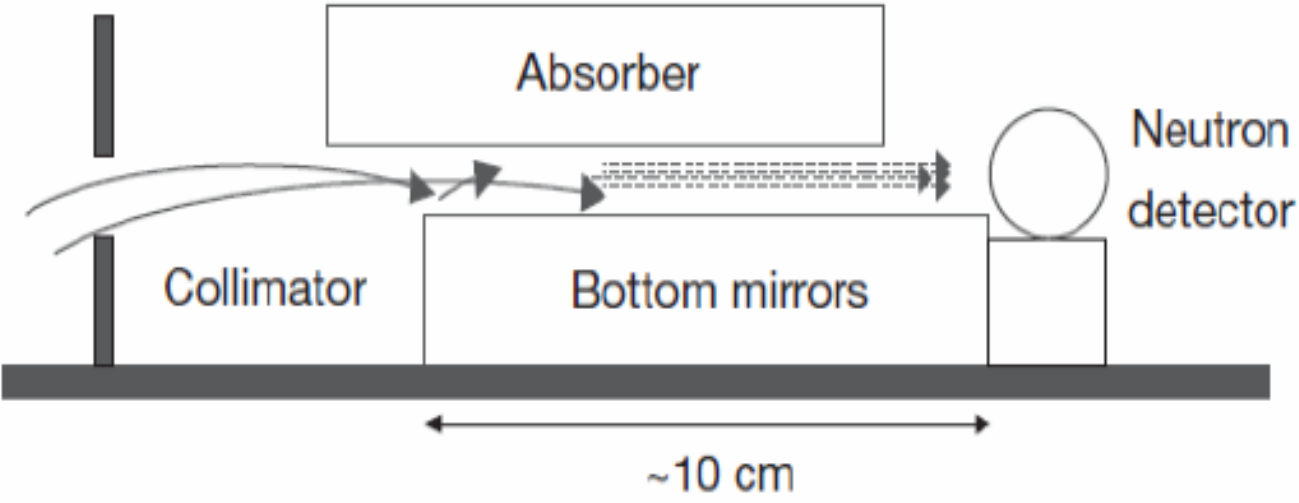}  \\       
         \includegraphics[width=13cm]{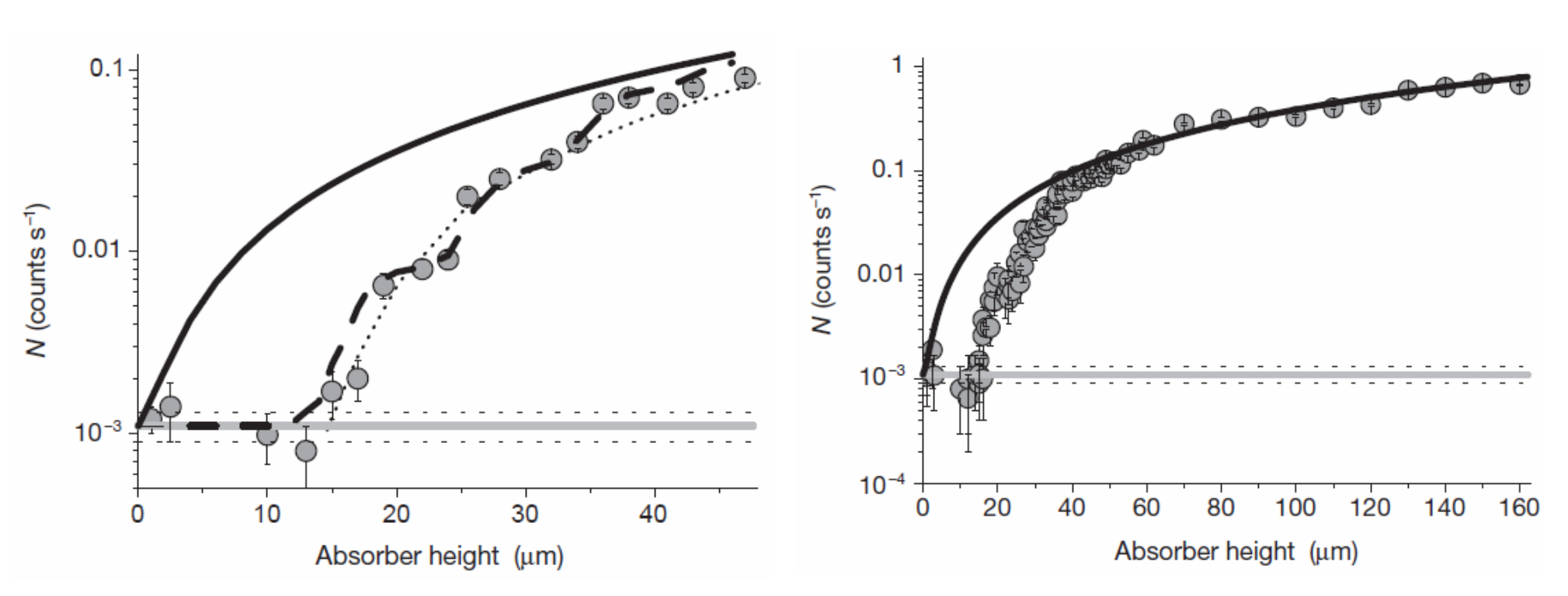}
         \caption{Slit experiment with ultra-cold neutrons. Upper picture: Experimental setup depicting the bottom mirror and the absorber, defining vertical boundary conditions. Neutrons enter through a vertical collimator from the left and traverse the vertical slit with variable absorber height. Traversing neutrons are counted without position sensitivity. Lower picture: Transmission curve for neutrons as a function of the vertical absorber position. The right graph shows the full vertical slit sizes explored depicting the asymptotic classical behavior. The left graph shows the transmission for the first 40 $\mu$\rm{m} exhibiting a step-like behavior for the transmission. Note that there is zero transmission for slit sizes below 16 $\mu$\rm{m}.}\label{UCN_transmission}
      \end{figure}

In the first experiment performed by \cite{Nesvishevski} the transmission of neutrons through a vertical slit was measured as a function of the slit height $h_{{\rm abs}}$. The slit was sufficiently long as to remove all neutrons with vertical velocities above $V_{{\rm grav}} = m\cdot g\cdot h_{{\rm abs}}$. The transmission curve reveals two striking behaviours. Although neutrons are quasi-point-like objects (on the scale of the slit size), they do not pass for values of $h_{{\rm abs}} \leq 15\mu \rm m$, revealing their wave nature. At large values of $h_{{\rm abs}}$ the transmission curve shows a step-like behavior with the position of the steps corresponding to the maxima of $\left|{\psi_n(h)}\right|^2$ for n=1, 2 and 3 (see fig.\ref{UCN_transmission}).

\par
\begin{figure}\centering
         \includegraphics[width=15cm]{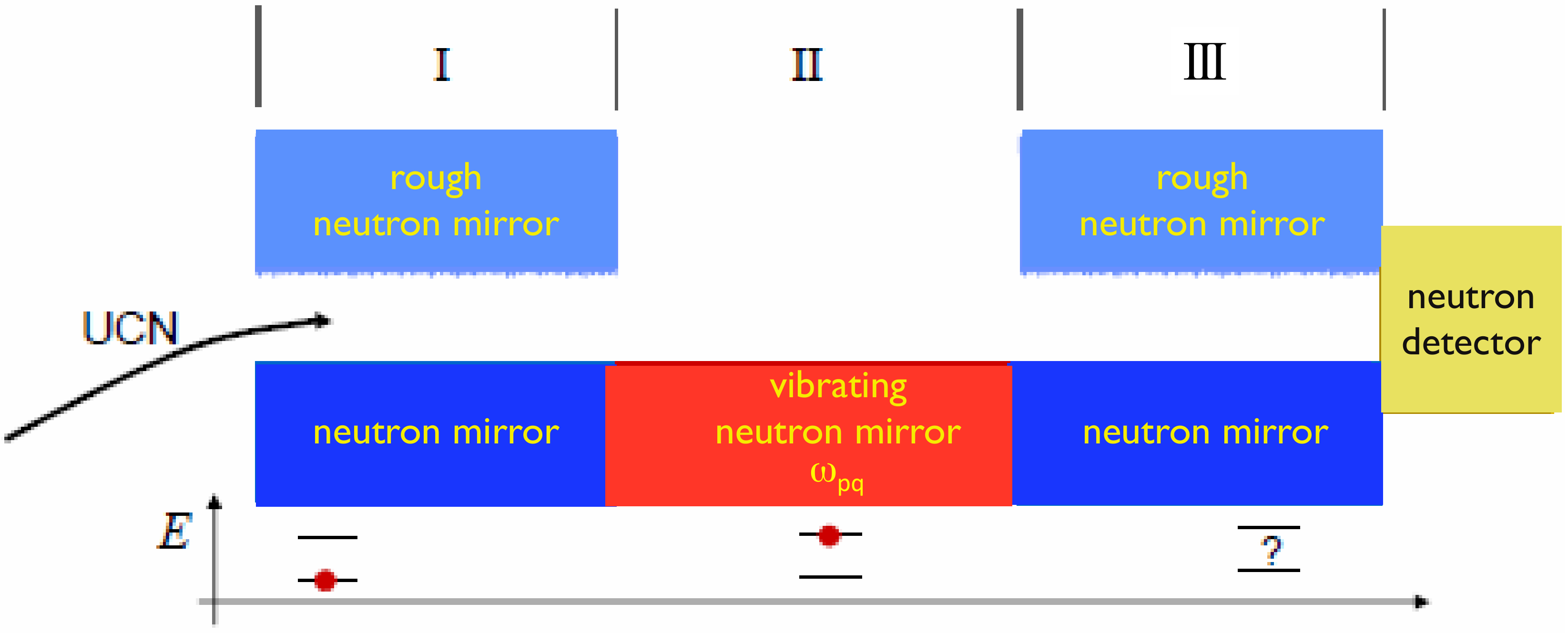}
         \caption{Neutron gravitational transition spectroscopy: UCN entering section I of the apparatus are prepared in the quantum state |1>. In section II resonance transitions are induced into state |x> and the neutrons remaining in |1> (section III) are counted by the neutron detector.}
         \label{resonance_spectroscopy_apparatus}
      \end{figure}

In order to probe the gravitational potential with higher precision and to omit the sensitivity to absolute distance and position measurements, Abele {\it et al.} have developed the technique of resonance transition spectroscopy for gravitationally bound neutrons \cite{Jenke11}. The setup, shown in fig.\ref{resonance_spectroscopy_apparatus}, consists of three sections in principle. A small vertical slit is used to prepare neutrons in the first quantum state, $n$=1. This slit is followed by a flat mirror path followed by a second slit, acting as analyzer with respect to the quantum states. The flat mirror can be put into vibration transferring energy to the neutrons being reflected from its surface and thus transitions from $n$=1 to any other value of $n$ can be induced.  

\begin{figure}\centering  
         \includegraphics[width=15cm]{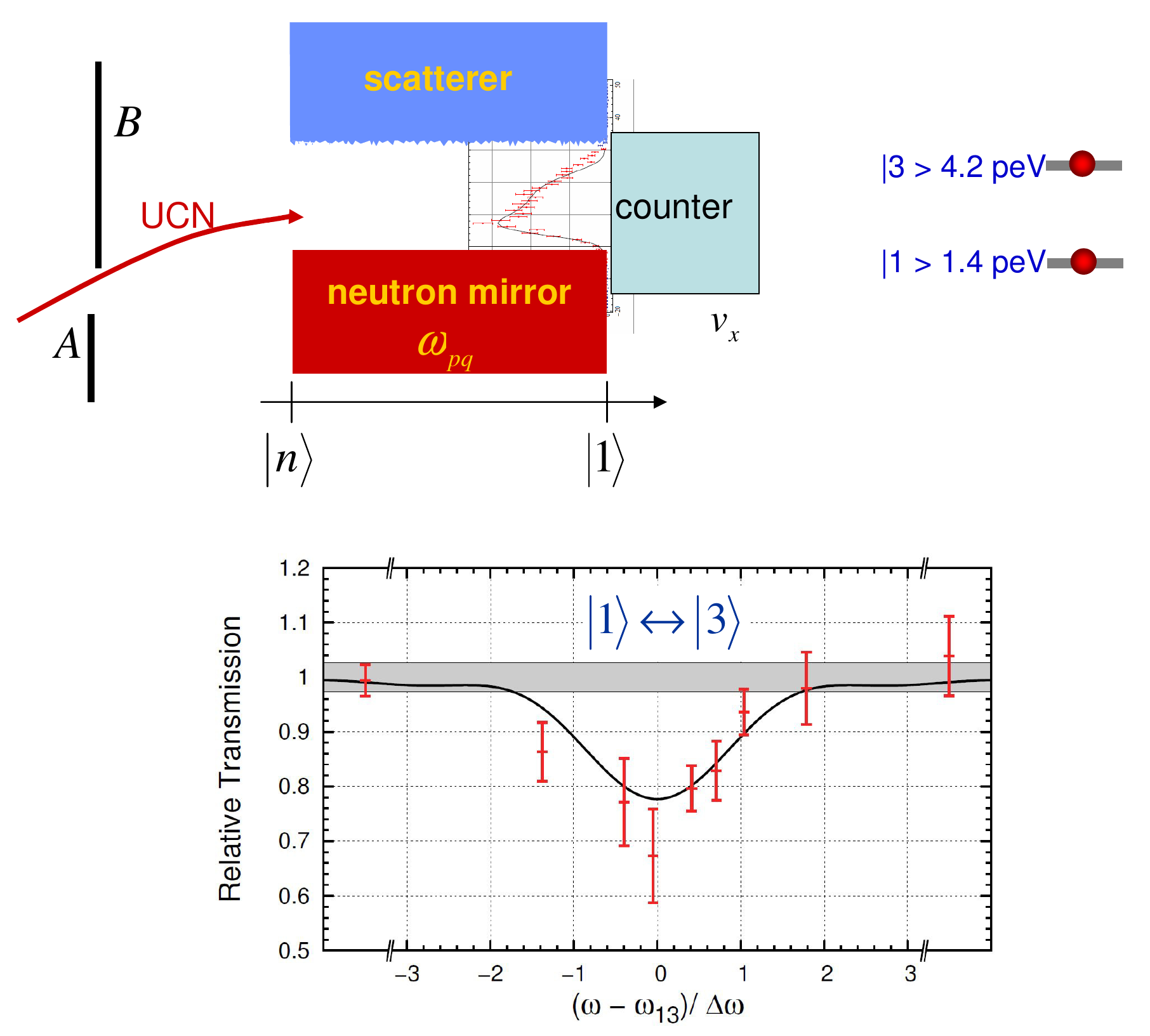}
         \caption{Upper picture: neutrons enter a slit of fixed height consisting of a bottom mirror and an absorbing lid, through a vertical collimator and are counted by a neutron detector at the right side. The bottom mirror can vibrate at frequency $\omega$ allowing energy to be transferred to neutrons on their path through the slit. Lower picture: transmission of neutrons as a function of the mirror vibration frequency $\omega$, here parametrized as the relative frequency deviation from the expected transition frequency $\omega_{13}$. The slight height had been optimized for the transmission of ground state |1> neutrons \cite{Jenke11}}
         \label{neutron_gravitational_resonance}
      \end{figure}

Fig.\ref{neutron_gravitational_resonance} shows a sketch of the setup used. For simplicity, there is only one slit which only transmits ground state neutrons. If the ground state is emptied due to resonance transitions, the transmission curve should change. Fig.\ref{neutron_gravitational_resonance}  also depicts the resonance curve, where the transmission is depicted as a function of oscillator frequency (at fixed oscillator amplitude) revealing clear sign of a resonance transition at the frequency expected. It corresponds to transitions from $n=1 \rightarrow n=3$. The experiment has also measured other transition frequencies. 

New developments are in progress \cite{Abele10} aiming at performing the experiments in the time domain, which follows the idea of N. Ramsey using split oscillatory fields, a very powerful technique also used for precision experiments described in the following section. Another experiment plans to perform resonance transitions by means of magnetic fields with a subsequent measurement of the neutron energy by means of a gravitational spectrometer (flight parabola) \cite{granit}. 
\par
\begin{figure}\centering
         \includegraphics[width=10cm]{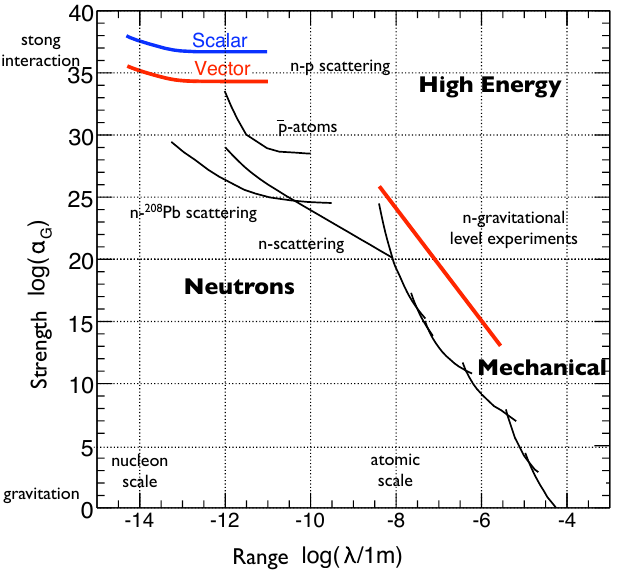}
         \caption{Limits for new light particles and $5^{th}$-force parametrized in terms of strength $a$ and range $\lambda$ of a Yukawa-like coupling. Taken from \cite{Kamyshkov2008} with additions from \cite{Dubbers}. Strength parameters for gravitational and strong force are indicated.}
         \label{neutron_gravitational_limit}
      \end{figure}

So far the experimental findings are in full agreement with the calculations made on the basis of Newtons's Law down to $\mu$\rm{m} distances. However, additional extra dimensions could manifest themselves just as a quasi $5^{th}$-force and mediated by an exchange boson which can be described by a Yukawa-type correction to the gravitational potential ($a\cdot e^{-r/\lambda}$). This exchange boson could be identified as an exchange of Kaluza-Klein particles (see e.g. \cite{Burgess2000}) for distances ${r}$ larger than the size of the dimensions. So far the measurements put only upper limits on strength {\it a} and range  {\it $\lambda$} of the interaction being equal to the Compton wavelength of the boson exchanged. \\
Fig.\ref{neutron_gravitational_limit} depicts the current situation in the plane of strength and range. The curves depicted give upper limits with the area above the curves being excluded by various experiments. For ranges around 1 $\mu$\rm{m} neutrons are an excellent tool with, at present, no principle limitations owing to the neutrality of the particle and extremely small polarizability. Thus, further improvements within the next five years are expected and can be extended to spin dependent forces using polarized neutrons.

\section{Search for a Neutron EDM}
Electric Dipole moments are a unique observable sensitive to new physics active at very small distance scales (and thus high energy scales). In fact, experiments putting upper limits on the neutron EDM have excluded more new theories in particle physics than any other system and offer more science potential, in particular as the scale for new physics has been pushed above 1 TeV within the first years of running at LHC. However, further improvements for the sensitivity for electric dipole moments must be connected to new developments in the field of sources for ultra cold neutrons, magnetic shielding, magnetometry and production of high electric fields. Owing to their high physics potential EDM experiments are planned or being setup by many groups at various locations, always close to strong neutron sources, such as reactors or spallation sources.\\
The action of an electromagnetic field on a neutron possessing a magnetic and electric dipole moment  $\mu_n$ and $d_n$, respectively, is given by: 
\begin{equation}
H_{int} = (\vec{B}\cdot\vec{\mu_n} + \vec{E}\cdot\vec{d_n})
\end{equation}
Owing to the different transformation laws for the electric and magnetic fields and the dipole moments under time reversal (T), charge conjugation (C)  and parity (P) leads to $H_{int}$ is not invariant under T-transformation. If we assume physics to be invariant under CPT, the presence of both terms in $H_{int}$ also breaks CP-invariance. Note that $\vec{\mu_n}$ and $\vec{d_n}$ should point along the directon of the spin $\vec{S}$ of the neutron.\\
Suggestions for a microscopic origin of $d_n$ are manifold, as it could originate from a CP violating part in QCD and thus being related to the so called $\theta$-term in the Lagrangian or from super-symmetry with its large parameter space. The latter is often quoted to set a scale between the present limit of $d_n\leq 2.6\cdot 10^{-26}{\rm e\cdot cm}$ \cite{EDM_RAL} and $d_n\geq 10^{-28}{\rm e\cdot cm}$. An electric dipole moment for neutrons originating from the Standard Model is at a scale of $d_n\approx 10^{-32}{\rm e\cdot cm}$ \cite{SM_EDM}, and thus presently untedectably small.
\par

\begin{figure}\begin{centering}
         \includegraphics[width=5cm]{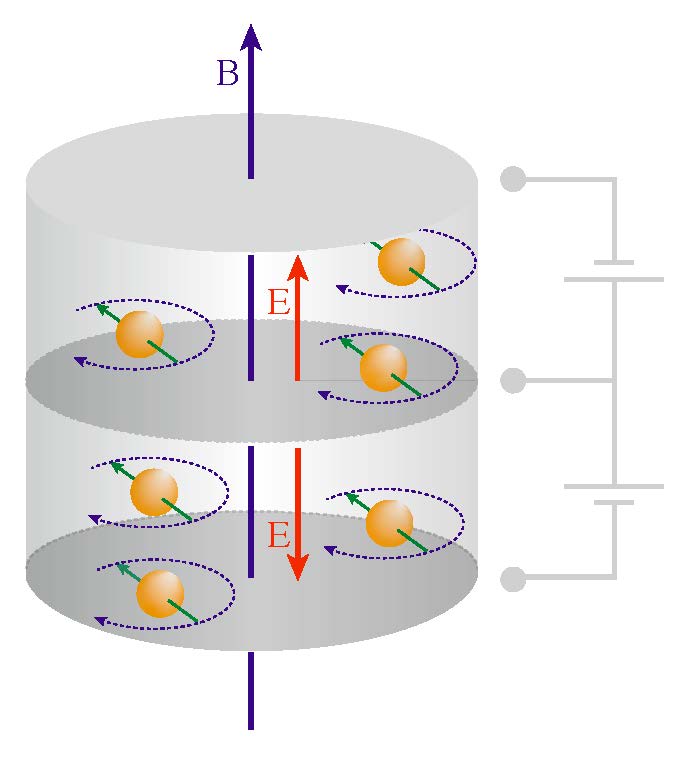}\\
         \caption{Principle of measuring an electric dipole moment of the neutron. Neutrons, stored in a box with axial magnetic field perform Larmor precession which can be altered in the presence of an electric field. If stored in separate boxes with opposite electric field direction, the comparison of precession frequencies in up and down cells directly yields the interaction strength, with the electric field as the magnetic field effects cancel in the two boxes}\label{sketch_EDM_search}\end{centering}
      \end{figure}

 Experimentally, electric dipole moments are searched for using the magnetic dipole moment as a reference. Ultra cold neutrons are stored inside a bottle with a superimposed magnetic field which leads to a spin precession of the magnetic moments (Larmor precession of frequency 30$\rm Hz$ in a typical magnetic field of 1 $\mu\rm T$) (see fig.\ref{sketch_EDM_search}). The storage volume is now subdivided into two volumes with electric fields superimposed parallel and anti-parallel to the magnetic field, respectively. The change in the neutron precession frequency now is proportional to the size of $d_n$ and is opposite in the two cells. The difference of the precession frequencies in the two cells is given by:
\begin{equation}
\hbar\cdot\Delta\omega = \hbar\cdot(\omega_{\mathrm{Larmor}}^{\Uparrow\Uparrow} - \omega_{\mathrm{Larmor}}^{\Uparrow\Downarrow}) = 4\cdot E\cdot {d_n}
\end{equation}
Thus, the measurement of an interaction energy is now translated to the measurement of a frequency difference.

\begin{figure}
       \centering
	\includegraphics[width=0.45\textwidth]{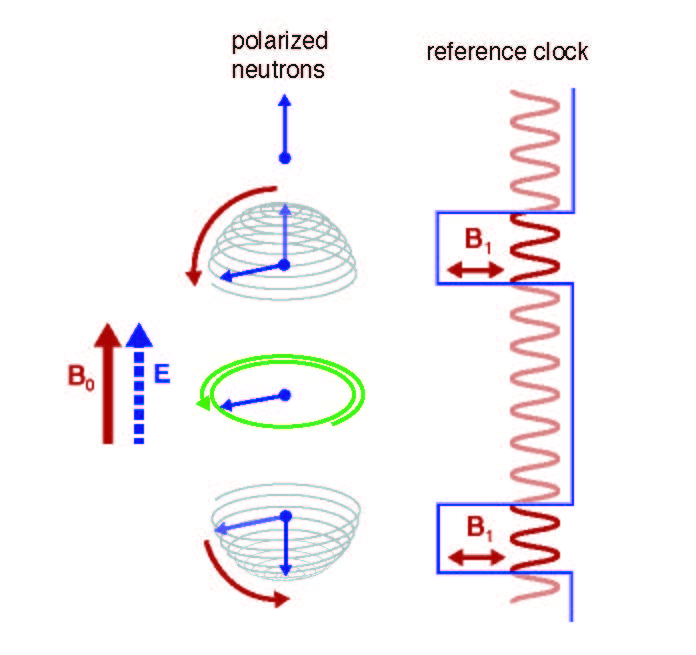}
	\includegraphics[width=0.45\textwidth]{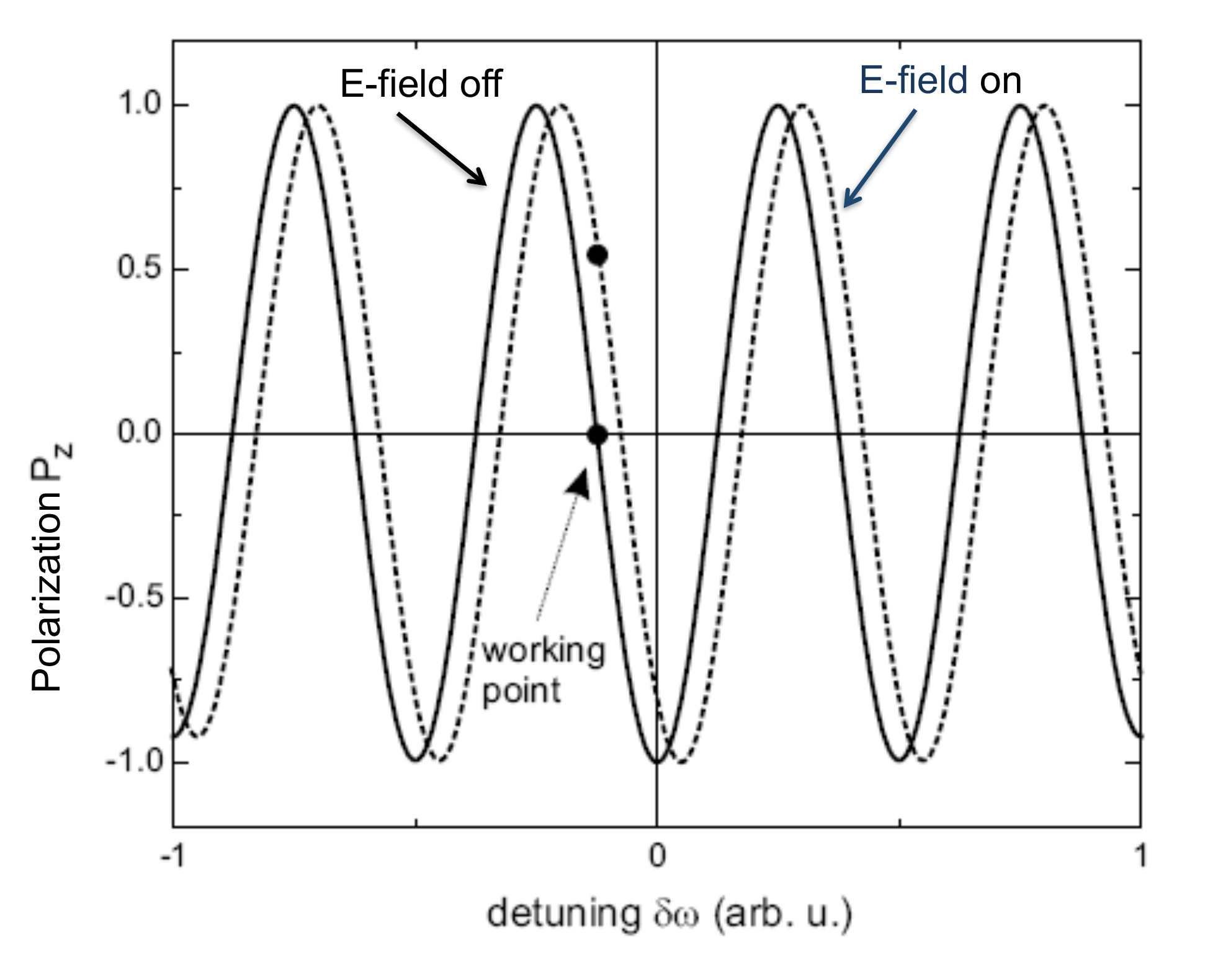}
         \caption
         {Principle of  Ramsey's method of separated oscillatory fields \cite{ramsey57}: Left: two clocks, an external reference oscillator and the neutron precessing in a magnetic field, are synchronized by an RF pulse which causes a $\pi/2$ flip of the spin into a plane perpendicular to the B-field. Both clocks are left running independently with the neutron being exposed to external fields (B-field alone or B and E fields superimposed).  By means of a second RF pulse the phases of the two clocks are analyzed as the efficiency for spin flip reversal is strongly dependent on it. The frequency of the external oscillator should correspond to $\omega_{\mathrm{Larmor}}$. Right: axial polarization $P_z$ of neutrons after the second RF pulse as a function of the phase shift (Ramsey pattern). The working point, corresponding to the electric field being switched off, is chosen to lay on the zero-crossing, as $dP_z/d\delta\omega$ is highest.  The dashed curve depicts the Ramsey pattern in the presence of an EDM, which shifts the resonance curve in case of the E-field being turned on between the two RF pulses.}
	\label{Ramsey}
      \end{figure}

\begin{figure}\begin{centering}
         \includegraphics[width=12cm]{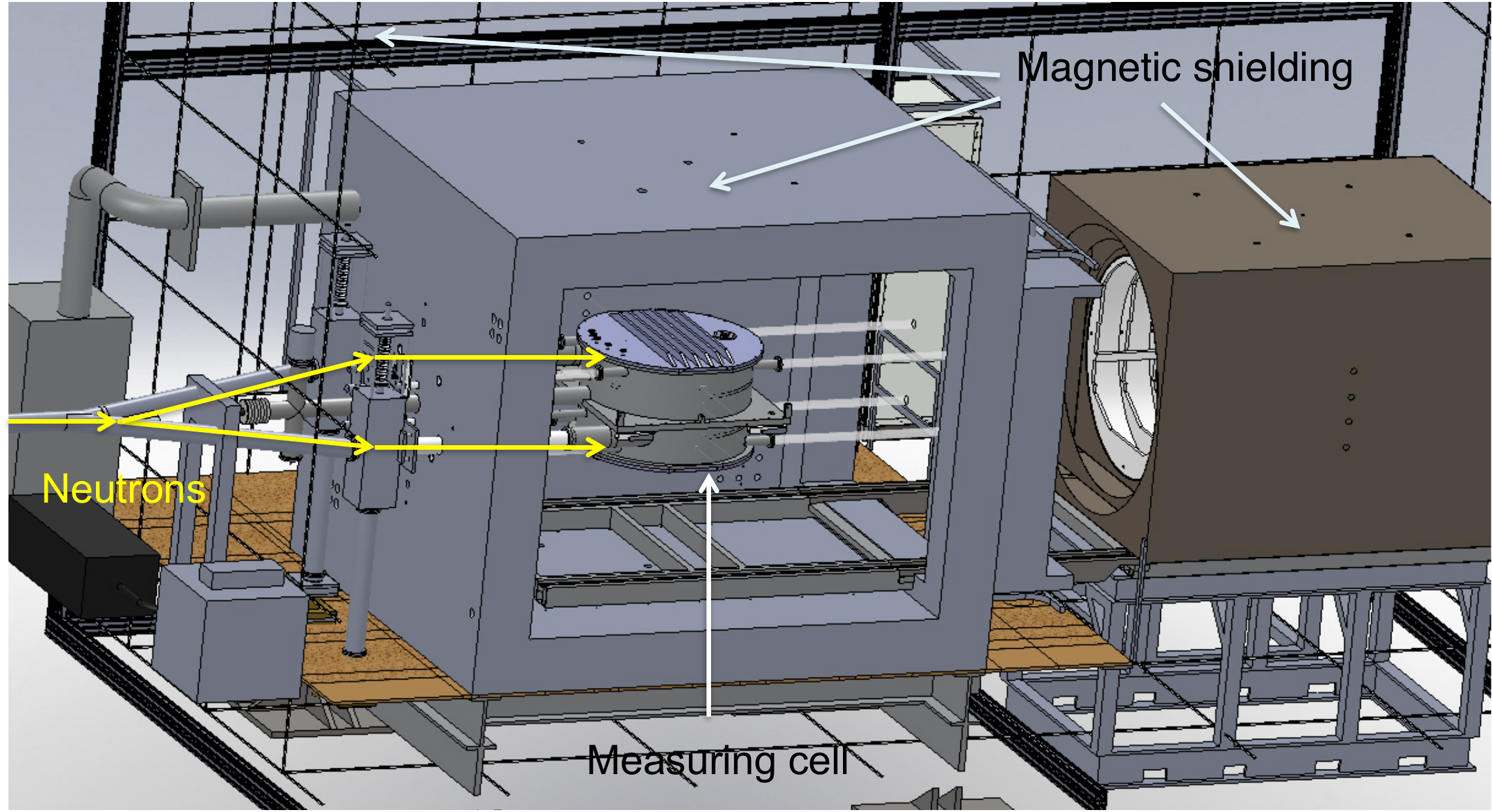}\\
         \caption{Sketch of the setup planned in Munich to measure the neutron EDM. An external field cage and two independent $\mu$-metal housings with very high magnetic shielding factors are shown. The EDM-measuring device consist of two storage cells, superimposed vertically with opposite electric field directions and placed in the centre of the shielding system. The path for polarized into and from the measuring cells is depicted on the left side of the shield \cite{fierlinger2012}. }\label{nEDM_MUC}\end{centering}
      \end{figure}

The most elaborate method to detect an EDM is based on the above technique and  still uses Ramsey's scheme of separated oscillatory fields \cite{ramsey57}, as sketched in fig.\ref{Ramsey}. Polarized neutrons are stored in a box with a superimposed magnetic dipole field ($B_0$-field) of $O(\mu \rm T)$ leading to $\omega_{\mathrm{Larmor}}\sim 30 \rm Hz$. By means of an external RF field controlled by a clock the neutron spin is oriented perpendicular to the magnetic field ($\pi/2$-flip) and thus begins Larmor precession. The neutron spin can thereby acts as neutron clock, the stability of which is only limited by the stability of the magnetic environment. After a typical exposure time of 100-200 seconds the neutron spin is again exposed to the RF field and thus flipped along the magnetic field direction. This can be regarded as a phase comparison between the neutron clock and the external clock as the efficiency of the second $\pi/2$-flip is directly related to the frequency difference of the two clock systems which is presumed to be very small. By varying the external clock frequency the polarization of the neutron with respect to the $B_0$-field direction undergoes Ramsey oscillations (see fig.\ref{Ramsey}) and the sensitivity to modifications of the Larmor precession is largest at the zero-crossing of the Ramsey curve exhibiting the largest slope. Now superimposing an electric field adds sensitivity to $d_n$ which will shift the Ramsey curve. Thus, a correlation of such a shift with respect to the size of the external electric field constitutes a direct measure of $d_n$.\\
Unfortunately the effects are small and the sensitivity currently obtainable corresponds to hypothetically controlling the spin precession of a single neutron to one turn in half a year. This is systematically limited by the control of the magnetic environment. Small field fluctuations or drifts will change $\omega_{\mathrm{Larmor}}$ in a random way, thus acting as noise. In addition, drifts of the magnetic field in time will lead to a shift of  $\omega_{\mathrm{Larmor}}$. This can in principle be controlled using reference systems made from polarized atoms like $^{3}\mathrm{He}$ or $^{199}\mathrm{Hg}$, the precession frequency of which can be monitored by  SQUID readouts placed above the magnetometer cells. These magnetometers can be installed around the neutron storage cell but within the same magnetic volume. \\
However, strong electric fields E, placed over the neutron storage volume can lead to surface currents in the walls of the storage cell which in turn are the origin of magnetic fields proportional to E, thus mimicking an EDM. In order to control such effects, $^{199}\mathrm{Hg}$-atoms are filled into the same storage cell as the neutrons (a co-magnetometer) and thus are exposed to the same field distortions. The $^{199}\mathrm{Hg}$ spin precession of $\sim$ 8Hz is monitored by polarized laser light shining perpendicular to the magnetic field axis, the absorption of which is time dependent and directly related to $\omega_{\mathrm{Larmor}}^{\mathrm{Hg}}$. However, owing to the mass difference of neutrons and $^{199}\mathrm{Hg}$-atoms,  co-magnetometer atoms probe a different volume of the storage bottle due to sagging in the gravitational field. This can be a problem if the magnetic field exhibits field gradients leading to non-uniform radial magnetic fields over the volume of the storage bottle and which depend on the position along the symmetry axis of the storage cells. Overall, field stabilities and homogeneities in the order of $10\mathrm{fT}$ and $0.3\mathrm{nT/m}$ must be achieved.\\
In the presence of an electric field these field gradients lead to geometric phases again proportional to E. The control of such subtle effects require excellent mapping and control of the magnetic field, for which reason great efforts are undertaken to improve magnetic shielding and magnetometry.\\
Once this problem is brought under control by a suitable setup the statistical accuracy of these measurements is related to the total number of stored neutrons, $N$,  the exposure time, $T$, the electric field of strength, $E$ and the visibility of the Ramsey-fringes, $\alpha$, which is in turn related to the spin polarization and spin analysis as well as the spin coherence time. We can thus define a figure of merit $\mathcal{M}$:
\begin{equation}
\mathcal{M} = \alpha\cdot\sqrt{N}\cdot T\cdot E
\end{equation}
As an example fig.\ref{nEDM_MUC} shows a setup proposed for the FRMII in Munich. Spin polarized ultra cold neutrons from a solid deuterium UCN source at the FRMII (envisaged to deliver UCN densities  $>3\cdot 10^{3} {\rm cm^{-3}}$ into the experiment starting at the end of 2013 \cite{Frei2007}) enter a two-cell storage system placed inside a magnetic shielding system. The latter consists of a magnetically shielded room with an active compensation system made from 24 coils and controlled by 180 magnetic field probes and an independent inner shielding cage. A specially designed magnet with so called $\cos{\theta}$-coils provides a very homogeneous magnetic $B_0$-field of $\sim$$\mu \mathrm{T}$. As described above, the electric field of $\sim18 \mathrm{kV/cm}$ has opposite directions in the two storage cells and is aligned with the magnetic field direction. Initially polarized along $B_0$, an RF pulse causes a $\pi/2$ spin rotation starting the free precession of stored UCN. After a storage time of $\sim$250 seconds,  the second $\pi/2$-pulse turns the spin again into the direction of  $B_0$ and the storage cells are emptied with subsequent polarization analysis and neutron counting. The analysis is performed separately for the two UCN storage cells. The procedure is then repeated many times.\\

The stability of the magnetic field environment on the storage cells is controlled by$^{199}\mathrm{Hg}$ read by a laser based optical system also present in the UCN storage cells as well as in two cells above and below the storage sandwich. The expected systematic effects should be pushed below $2\cdot 10^{-28}e\cdot cm$ which should allow sensitivities for a n-EDM of $d_n < 5\cdot 10^{-28}e\cdot cm$ within 200 days of measuring time.\\
Similar experiments are being setup at PSI \cite{EDM_PSI} and TRIUMF \cite{TRIUMF} while so called cryogenic EDM-experiments, combining UCN production and EDM measurements within one setup are being  operated or prepared at ILL \cite{EDM_ILL} and SNS \cite{EDM_SNS} .

\section{Lifetime of the neutron}
\begin{figure}\begin{centering}
         \includegraphics[width=7.5cm]{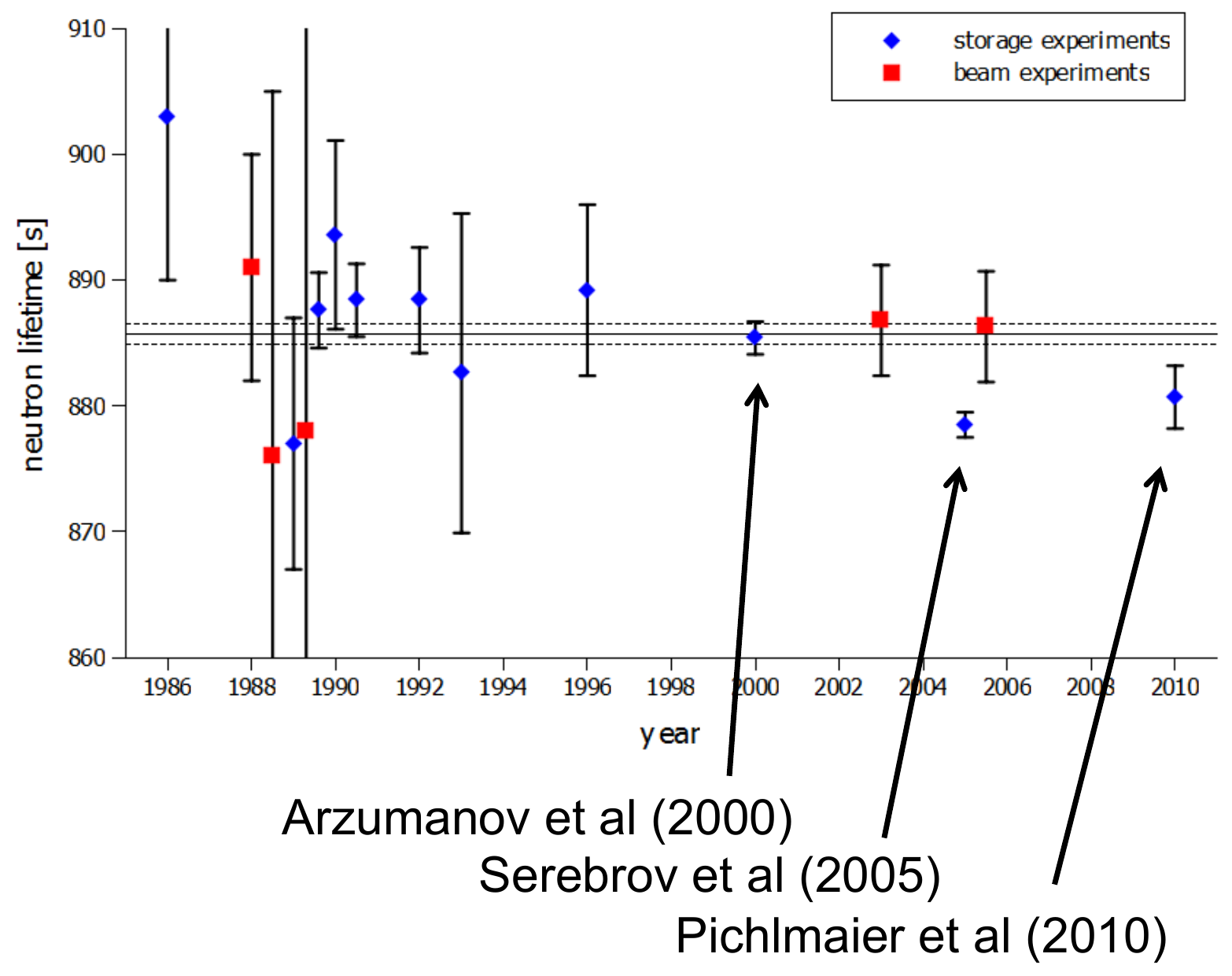}\\
         \caption{ Overview on the measured values for the neutron lifetime stretched along the time axis. As can be seen, the present situation exhibits discrepancies far beyond the individual measuring uncertainties quoted \cite{Arzumanov}, \cite{serebrov_1}, \cite{Pichlmaier_10}. The most precise measurements are quoted The solid line corresponds to the world average quoted by the particle data group \cite{PDG10}.}\label{lifetime_overview}\end{centering}
      \end{figure}

Despite more than half a century of measurements the lifetime of the neutron still remains a poorly known quantity with the two statistically most precise measurements differing by more than six standard deviations, corresponding to about 8 seconds (see fig.\ref{lifetime_overview}). This clearly points towards an uncertainty with respect to systematic effects, but owing to low count rates only limited studies of these could be performed and may in part be inherent to the measurement methods. Neutron decay thereby has relevance to different fields as briefly outlined below.
\subsection{Physics related to the neutron lifetime}
\begin{figure}
       \centering
         \includegraphics[width=0.55\textwidth]{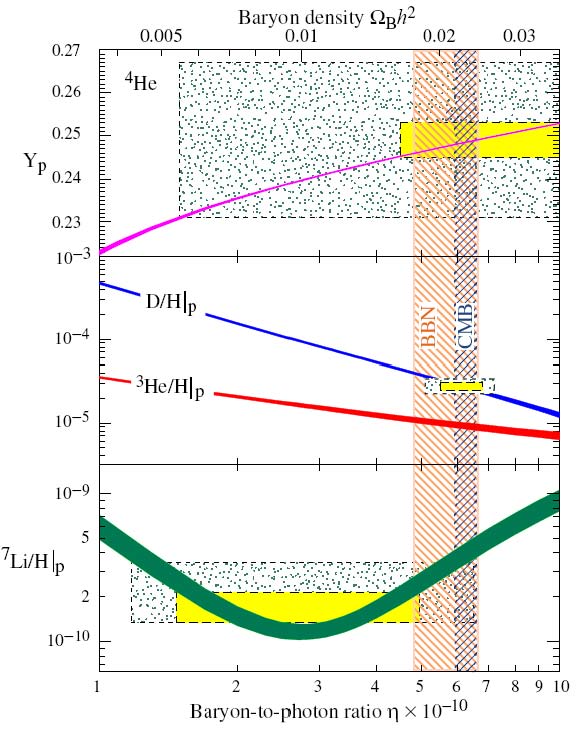}
         \includegraphics[width=0.35\textwidth]{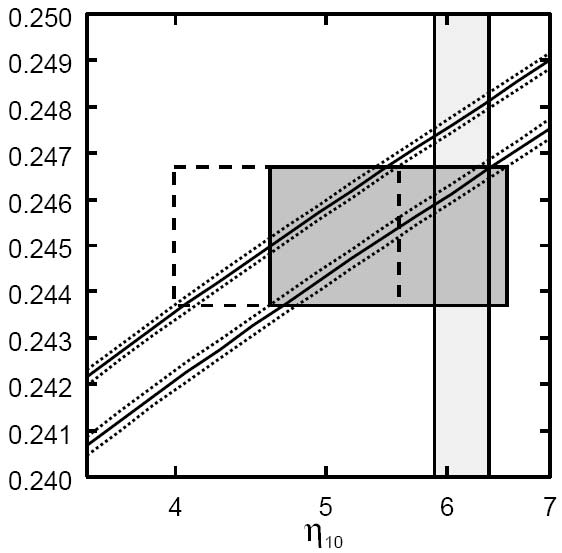}
         \caption{Left: Isotope mass fraction versus the baryon fraction in the Universe using the standard model of cosmology (lines). The solid areas depict the statistical errors for astronomical observations, the dotted ones the systematic uncertainties \cite{BBNS}. Right: Helium mass fraction versus the baryon fraction in the early Universe for two different values of the neutron lifetime \cite{BBNS_2}. Upper: PDG value \cite{PDG06}, lower: ref. \cite{serebrov_1}. The boxes show the allowed values for the baryon mass fraction with the box size indicating statistical uncertainties for $Y_p$ only. The vertical band shows the baryon fraction deduced from WMAP data.}
         \label{helium_abundance}
      \end{figure}

One of the key processes with relevance to neutron decay is primordial nucleosynthesis \cite{BBNS}. A few minutes after the Big Bang , the weak interaction caused an equilibrium of neutrons and protons owing to the reactions $n\rightarrow p e^-\overline{\nu}_e$ and the electron capture reactions $p e^-\leftrightarrow n\nu_e$ and $n e^+\leftrightarrow p\overline{\nu}_e$. The equilibrium of these reactions was broken once the expansion rate of the Universe won over the mean free path of the neutrinos (governed by the strength of the weak interaction $\Gamma_{n\leftrightarrow p} \sim G^2_{\mathrm{F}}\cdot T^5)$. At this temperature $T$ neutrinos decouple from the system and $T$ determines the $n/p$ ratio $n/p = e^{-Q/T}$,
where $Q=1.293~\rm{MeV}$ is the neutron-proton mass difference. This ratio changes subsequently owing to free neutron decay. As the Universe expanded the temperature droped below the photo-dissociation threshold for deuterons and efficient nucleosynthesis began, leading to the production of light elements like  deuterium, helium and lithium. The abundance predictions of the Standard Model of cosmology using the neutron lifetime as input parameter is shown in fig.\ref{helium_abundance} as function of the
baryon-to-photon ratio $\eta_{10}$, where $Y_{\mathrm{P}}$ denotes the helium mass
fraction in the early Universe \cite{BBNS}.
The left plot in fig.\ref{helium_abundance} demonstrates, as an example, the effect of changing the neutron lifetime in the model \cite{BBNS_2}. Although having big influence,
 the value of $Y_{\mathrm{P}}$ determined from the He-spectroscopy in low metalicity regions in our Universe is not yet measured with sufficient precision and systematic uncertainties in the extrapolation of the helium abundance to regions with zero metalicity dominate the
experimental error band. Thus, the consistency of the Standard Model is not yet in question.\\

\subsection{Particle Physics} In the Standard Model, neutron decay is governed by the weak interaction with
the underlying V--A structure. The Lagrangian contains two parts, a leptonic and a hadronic one. The latter is written as \cite{decay_formula}:
\begin{eqnarray}
\label{V_A}
    V_\mu-A_\mu & = & \emph{i}\overline{\Psi}_p\{f_1(q^2)\gamma_{\mu}+f_2(q^2)
    \frac{\sigma_{\mu\nu}q^{\nu}}{m_p}+\textit{i}f_3(q^2)\frac{q_{\mu}}{m_e}\}\Psi_n\nonumber\\
    && -\emph{i}\overline{\Psi}_p\{f_i\rightarrow g_i\gamma_5\}\Psi_n.
\end{eqnarray}
Using the conserved vector current (CVC) hypothesis most form factors $f_i$ and $g_i$ can be set to zero but
\begin{eqnarray}
\label{V_A_2}
    G_{\mathrm{V}}=f_1(q^2\rightarrow 0)\cdot V_{ud}\cdot G_{\mathrm{F}} = g_{\mathrm{V}}\cdot V_{ud}\cdot G_{\mathrm{F}}\nonumber\\
    G_{\mathrm{A}}=g_1(q^2\rightarrow 0)\cdot V_{ud}\cdot G_{\mathrm{F}} = g_{\mathrm{A}}\cdot V_{ud}\cdot G_{\mathrm{F}}\nonumber
\end{eqnarray}
and we obtain an expression for the first element of the Cabibbo-Kobayashi-Maskawa quark mixing matrix $V_{ud}$
\begin{equation}
\label{V_A_3}
    \mid V_{ud}\mid^2=\frac{1}{\tau_n}\frac{(4908.7\pm1.9)~{\rm s}}{(1+3\lambda^2)} ;~~~~~{\rm with}~~\lambda=\frac{}{}\frac{G_{\mathrm{A}}}{G_{\mathrm{V}}}=\frac{g_{\mathrm{A}}}{g_{\mathrm{V}}}.\nonumber
\end{equation}

The largest theoretical uncertainties come from radiative corrections which are common to both, free neutron decay and pure Fermi-transitions in nuclei \cite{hardy08}.
\subsection{Experimental perspectives}
In general, there are two classes of experiments, so called "in beam" experiments \cite{Paul08}, where an n-beam traverses a fiducial volume and the decay products are collected, and storage experiments, which are generally more accurate and principally self-calibrating. Here neutrons are stored in suitable bottles for a pre-defined storage time, $t_{\mathrm{storage}}$, and the remaining neutrons are counted while emptying the bottle. In the case of material bottles, which always exhibit wall losses the size of the box is varied and the resulting dependence on the box size is extrapolated to box size infinity, thus eliminating effects of losses due to storage. Systematics arise from e.g. keeping all experimental conditions constant while varying the box size. \\

       \begin{wrapfigure}{l}{6.5cm}
       	\vspace{-1cm}
                 \includegraphics[width=0.4\textwidth]{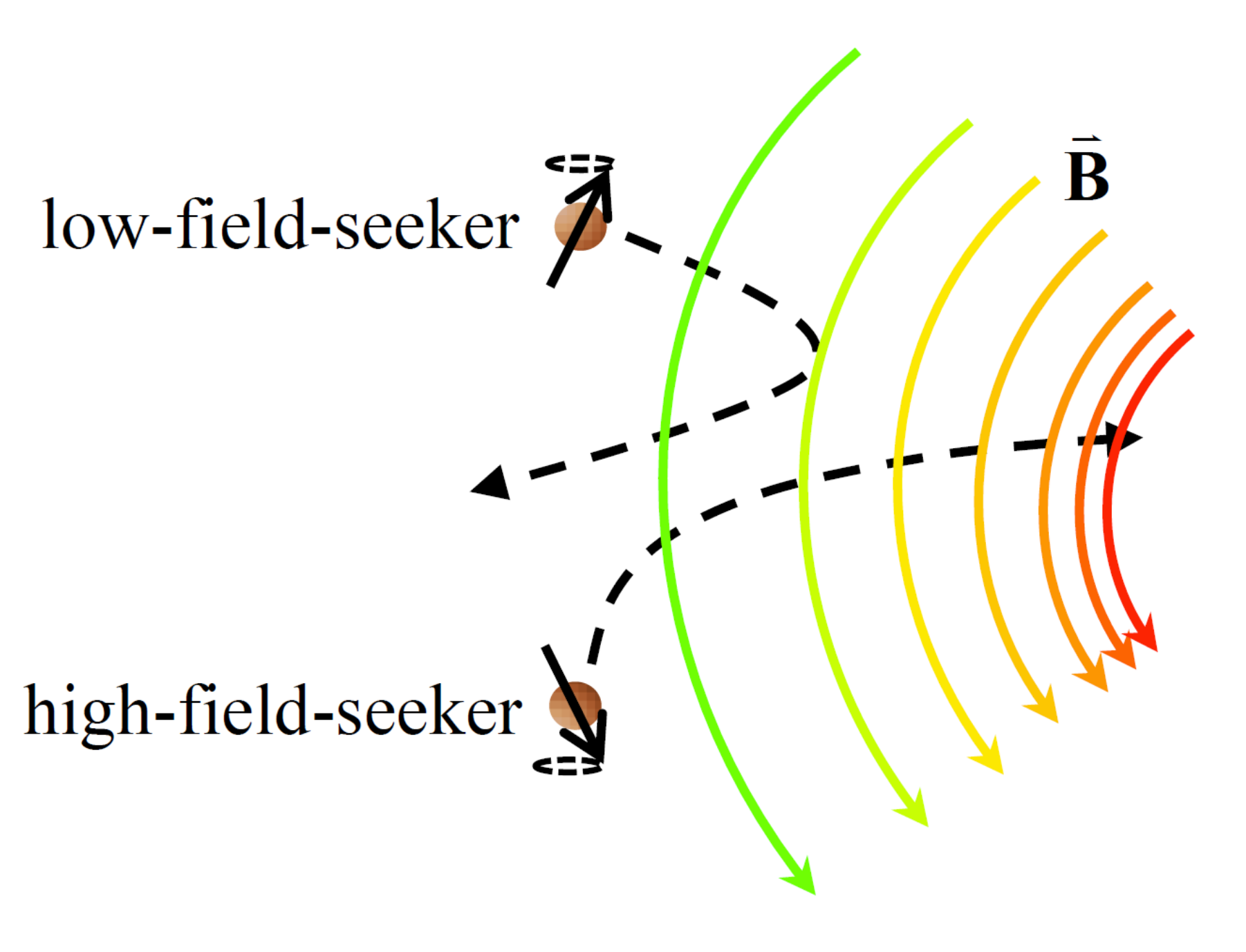}
        \vspace{-0.5cm}
                 \caption{Principle of low-field and high-field seekers.}

                  \label{field_seekers}
	\end{wrapfigure}
         
\begin{figure}\centering
         \includegraphics[width=0.35\textwidth]{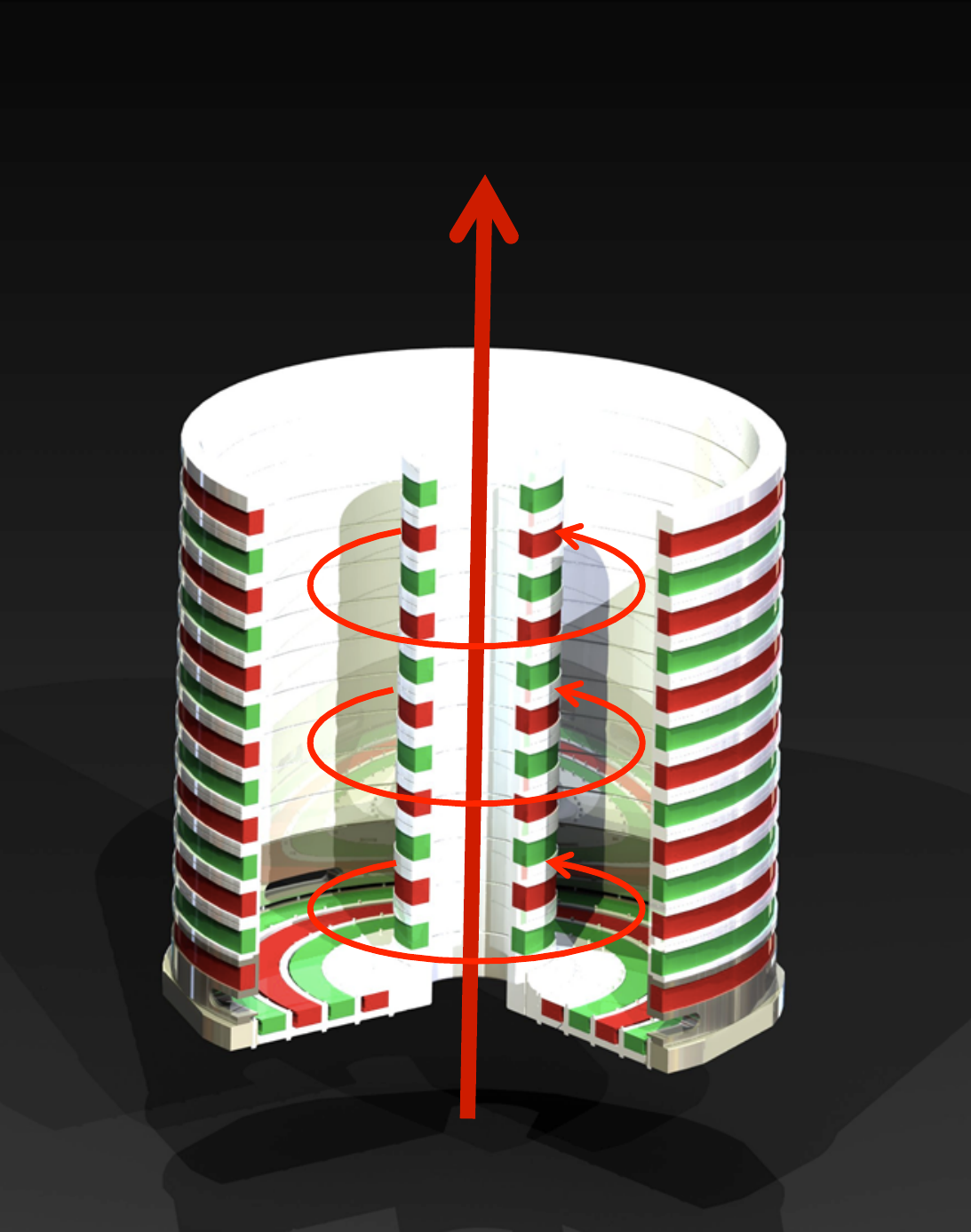}
                  \includegraphics[width=0.35\textwidth]{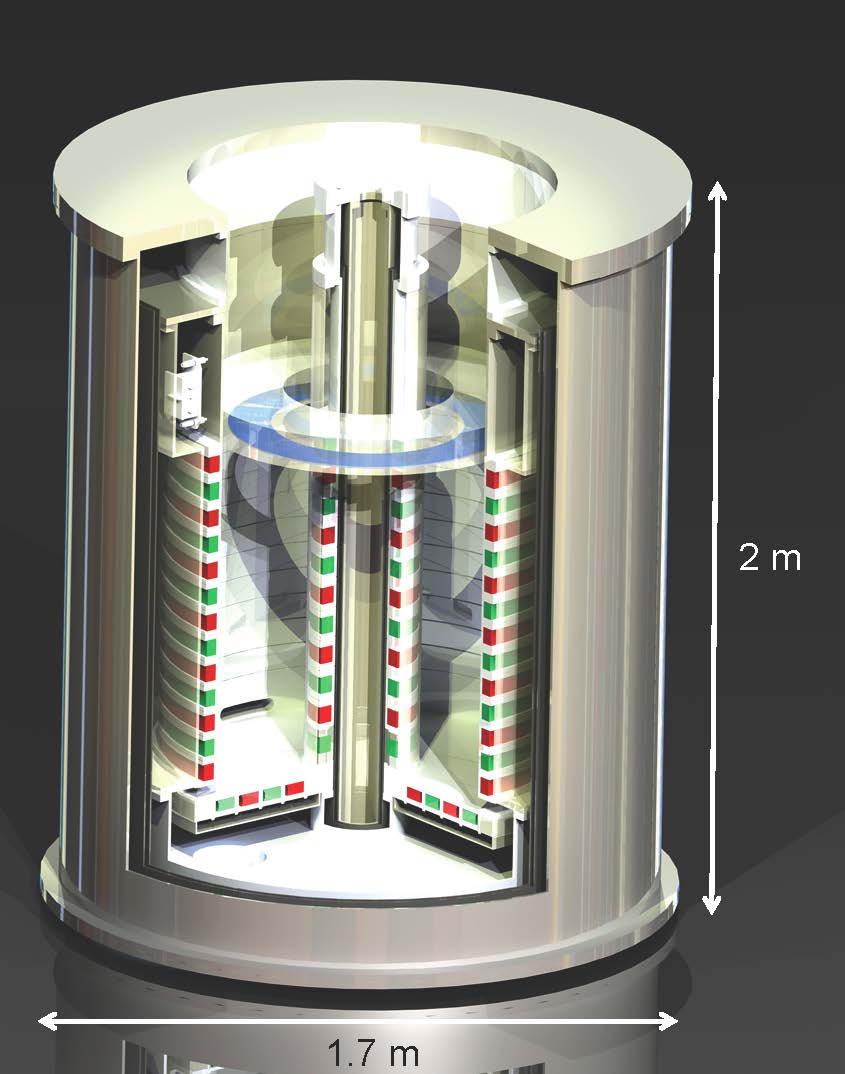}
         \caption{Sketch of magnetic trap to measure the neutron lifetime. A coil-system with the 28 inner and outer coils as well as the bottom coils operated with alternating current direction (coloured in red and green). The central current rods are also visible. Centre: coil-system implemented into the experimental setup with vacuum. The blue lid represents the proton detector system. Sizes indicated correspond to the vacuum vessel, the trap height is 120cm.}
         \label{Penelope_1}
      \end{figure}

\noindent An alternative approach is neutron storage by magnetic walls \cite{Picker}, which is loss-less if designed correctly. Polarized neutrons are attracted (repelled) by strong magnetic field gradients with 60 {\rm neV/T}, such that one speaks of high-field and low-field seekers, respectively (see right sketch in fig\ref{Penelope_1}). The potential energy $V_{\mathrm{mag}}$ is given by: 

\begin{equation}
\label{magnetic_storage}
V_{\mathrm{mag}} = \nabla\cdot (\vec{\mu_n}\cdot\vec{B})
\end{equation}

         \vspace{0.4cm}

Typical wall potentials are about 120 neV for fields around 2T. Thus, suitable magnetic multi-pole fields can be used to enclose neutrons (low-field seekers) in a bottle with open lids. As gravity acts on neutrons with similar strength, such bottles with height of 1.2m are able to store neutrons with kinetic energies up to 120 neV and only require side walls and a magnetic floor, just as coffee is kept in a cup. Care has to be taken to make local magnetic field changes slow enough as to obey the adiabatic conditions for spin transport without spin flips. Similarly, zero-field regions are to be avoided throughout the storage volume to avoid spurious losses again due to spin flip. One such design currently being realized is shown in fig.\ref{Penelope_1}. The multipole trap is built up from a stack of superconducting coils with alternating current direction. Such a configuration keeps neutrons inside a cylinder of height $\sim$1m. As this configuration alone would result in zero-field regions along the axis a cylindrical  magnetic field, generated by current rods along the central axis is superimposed. In order to avoid neutrons hitting the rods and being absorbed the central conductors are again shielded by a series of coils with alternating current, mirroring the field configuration of the outer coils. The key feature of this arrangement is the real time detection of decay products, namely slow protons which are accelerated towards a large area detector placed at the top of the bottle by means of a high electric field (typically 20--40 kV)\\
The measuring cycle comprises several steps. The filling process starts with the magnetic field turned off in order to allow UCN to enter from below into the bottle though filling slits. These neutrons are initially stored by material walls. The magnet is ramped up within 100 seconds driving the low field seekers away from the material wall and thus reducing the storage volume, thereby slowly heating the neutron gas. Remaining high field seekers can be trapped close to the reflecting surfaces. Their removal occurs by an absorber ring lowered into the storage volume from above leaving only low-field seekers and neutrons with velocities below a critical velocity $v_{store}^{max}$. After retraction of the ring the clock is set to zero and proton counting starts. The count rate should follow an exponential decay law from which the neutron lifetime is deduced. At the end of a storage cycle (about 300-3000 $s$) the magnetic field is ramped down and the remaining neutrons are extracted and counted for about 200 $s$. The experiment is then repeated $n$ times with different pre-selected storage times. The neutron lifetime can also be extracted from the distribution of the neutron population left after different pre-selected storage times, although with lower statistical accuracy. This method, however, has a different possible systematic bias.
       
The experiment is designed towards the study of systematic effects. With its large volume of $>500 l$ typically $10^7-10^8$ neutrons should be stored at a modern well-working UCN facility, allowing the determination of the neutron lifetime to 1$s$ accuracy after one measuring cycle of about 15 minutes. Thus, numerous cross checks are foreseen such as varying the storage potential, the filled UCN spectrum, the cleaning procedure after each filling process or the observation for spin-flipped neutrons. The final uncertainty for the lifetime should be below 0.1$s$.

Other experiments, mostly with a much smaller storage volume and/or fewer variations to test for systematic effects are also planned around the world. The most innovative approach is the combination of a neutron lifetime experiment with a UCN source based on superfluid helium \cite{Leung}.

\section{Addressing the neutrino helicity through the two-body decay of the neutron}
More than 50 years ago M. Goldhaber determined the helicity of $\nu_e$ to be $\mathcal{H}_{\nu} = -1$ with his pioneering experiment \cite{goldhaber} using a two step process following the decay of $^{152}\mathrm{Eu}$ which translates the helicity of the primary neutrino to the helicity of a secondary photon being emitted, which is directly experimentally observable. Since then the V-A theory of weak interaction has been developed as part of the Standard Model in particle physics with the lepton sector comprising only massless and left-handed neutrinos and right-handed anti-neutrinos. With the observation of neutrino oscillations it has emerged that neutrinos must possess mass, though possibly being very small and below the eV scale. Still, the question of the nature of the right-handed sector is discussed, both in particle physics and cosmology. The presently believed properties of the left-handed sector of neutrinos are related to a possible right handed sector by means of the seesaw mechanism in which the low mass of left-handed neutrinos is explained by super heavy right-handed neutrinos as partners, which in turn have been discussed as candidates for dark matter. \par
\begin{figure}\centering
         \includegraphics[width=0.9\textwidth]{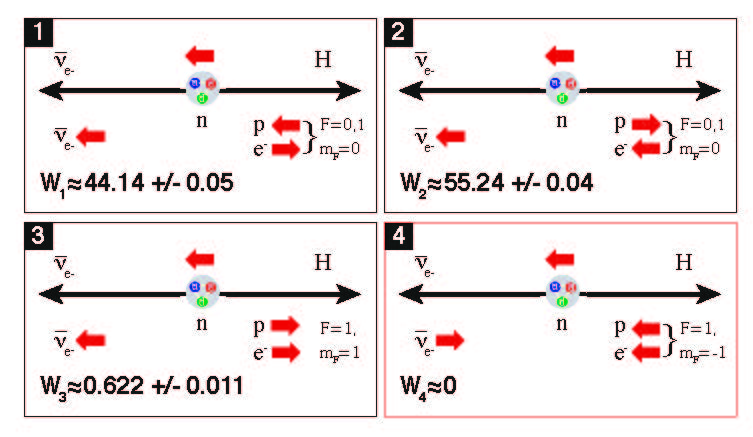}
         \caption{Spin correlations in neutron bound $\beta$-decay assuming the decay of a neutron with spin direction towards the left and the hydrogen being observed flying anti-parallel to this. The relative spin orientation of electron and proton make up the hyperfine quantum numbers of the $H$-atom ($F, m_F$). The upper configurations will give contributions to both, $F=0$ and $F=1$ and $m_F=0$ while the lower configurations only contribute to $F=1$. For right-handed anti-neutrinos configuration $4$ (lower right) is forbidden. The numbers indicate the population of the configurations assuming standard model $V-A$ coupling.}\label{bound_beta_1}
      \end{figure}

Currently the best measurement of $\mathcal{H}_{\nu}$ uses $\mu$ decay \cite{twist} where it enters the description of the electron spectrum typically analyzed in terms of Michel parameters \cite{michel}. Lower limits for $\mathcal{H}_{\nu}$ are now about 97\%. However, neutrons offer an alternative approach to address $\mathcal{H}_{\nu}$ as the phase space for normal neutron decay ($n\rightarrow p e^- \overline{\nu}_e$) offers a small corner for the  $p$ and $e^-$ to coalesce into a hydrogen atom, thus resulting in an effective two-body decay with an expected branching fraction of $\approx 10^{-6}$. Owing to the short range of the weak interaction only atomic S-states will be populated. In this two body decay not only the momenta of the H-atom and $\overline{\nu}_e$ are correlated ($p_{\overline{\nu}_e} = - p_H$) but also the spin correlation of $p$ and $e^-$ and the population of hyperfine states (HFS) reflect the neutrinos helicity as depicted in fig.\ref{bound_beta_1}.

 The population of the different HFS depends on the structure of the weak interaction. It can be altered by the presence of scalar or tensor interaction with strength $g_S$ and $g_T$. The forbidden HFS cannot be populated directly, if $\mathcal{H}_{\nu} = -1$ is exact. However, the population of $n$S-states with $n\geq 4$ will lead to a feed down populating $2$S-states according to the Einstein coefficients. For reasons discussed below, only direct observation of the $2$S state is interesting as it is both metastable and can be quenched to the ground state by different methods.This allows HFS analysis and discrimination against background hydrogen atoms. The population of the different states with principle quantum number $n$ can be calculated, the correction due to the feed down is known. It however spoils a "zero"-measurement as it constitutes a background.
So far, upper limits on $g_S$ and $g_T$ come from nuclear and neutron decay correlations and are rather poor. The study of the two-body decay would thus offer an alternative and in principle very sensitive way to measure these parameters.
 \par
\begin{figure}\centering
         \includegraphics[width=0.9\textwidth]{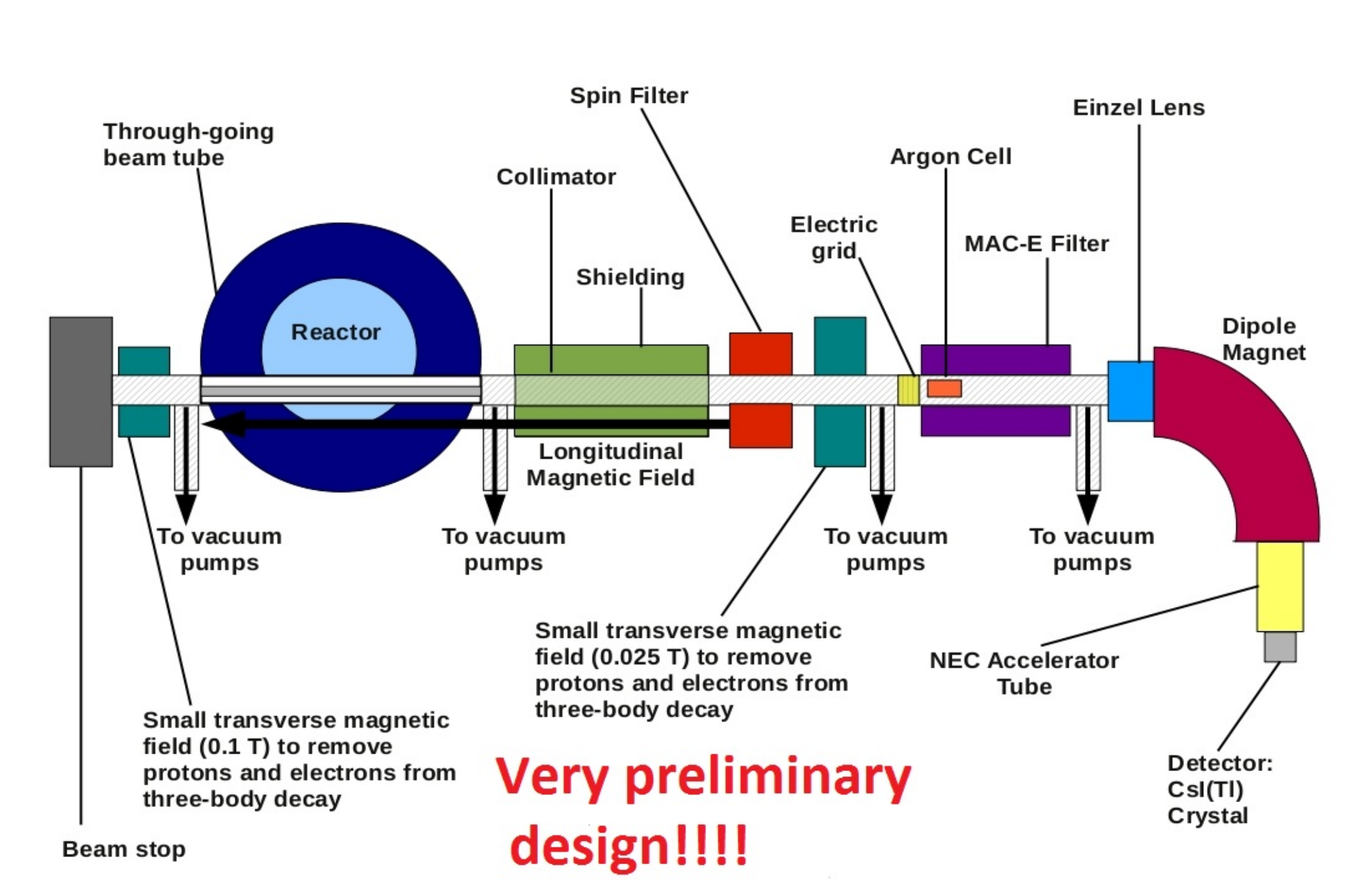}
         \caption{Schematic diagram of the layout of an experiment searching for the bound $\beta$-decay of the neutron at a research reactor. $H$-atoms emerge from a tangential beam tube acting as a neutron decay volume and $H$-extraction tube simultaneously. A set of collimators and shielding elements protect the measuring apparatus from a high flux of $\gamma$ and neutron background. Outside the biological shield the spin filter and $H$-analysis system are found. The latter uses a gas filled $Ar$-filled cell for charge-exchange (see text) followed by an energy selective magnetic energy analyzer filter and a magnetic spectrometer. $H^-$-ions are post-accelerated to about 20-40 keV and detected in a CsI-detector. At the rear end of the beam tube a beam stop in conjunction with a magnetic field deflecting charged $n$-decay particles is meant to reduce backscattering into the detector hemisphere.} \label{bob_layout}
      \end{figure}

Attractive at first sight with its virtue being discussed in several publications already 40-50 years ago \cite{Nemenov} \cite{Bahcall} \cite{Song} this decay has not yet been observed. Both the small branching fraction and the difficulties observing and identifying fast neutral hydrogen atoms constitute a challenge to experimental physics. A strong source of decaying neutrons must be combined with an atomic physics experiment. Due to the strong background of charged particles, neutrons and gammas produced by the neutron source certain requirements were imposed. Several research reactors possess tangential and through-going beam tubes which have the virtue of being traversed by a strong neutron flux but in turn offer no direct view to the reactor core. As the tube is evacuated, no neutrons should traverse it along its axis as no scattering process can alter the direction of the neutrons always entering the tube under a finite angle with respect to its axis.
Thus, only neutron decay products ($\rm p, e^-$) and gammas from neutron activation will travel along the axis, together with a very small fraction of hydrogen atoms from the two-body n-decay, only 10\% of which are in the $2$S-state. Using the Munich research reactor FRMII as an example, we expect about 0.3 usable hydrogen atom per second emerging from a tangential beam tube after about 10m of flight out of the biological shield arriving at random times \cite{Schott}. The sketch of a principle layout is shown in fig. \ref{bob_layout}.
\par
An experiment to study this decay will be performed, the first goal being the detection of the very rare decay itself, followed by the hyperfine analysis of the emerging hydrogen atoms. The principle challenge is the identification of the mono-energetic hydrogen atoms emerging from this decay with 326.5 eV kinetic energy. These hydrogen atoms must be separated from the large number of hydrogen atoms from the rest gas and from fast hydrogen atoms generated by protons from n-decay hitting the wall material at energies up to 730 eV.  This could be done using suitable detectors sensitive to neutral particles and having high energy resolution. Such detectors are not available at present. Thus, the identification must occur via ionization of the hydrogen atom. 

Resonant laser ionization exploring the Doppler shift would be the ideal way to ionize and detect the H(2S) atoms. However, this cannot proceed from the ground state owing to the absence of lasers in the far UV (10.6 eV). Laser ionization from the $2$S-state can only be done as a three step process, namely in a cascade $2S\rightarrow 10P\rightarrow 27D$ with subsequent field ionization. Owing to the large cross-section of the beam tube as well as the random arrival of hydrogen atoms a high-power DC laser system must be used. In addition, owing to the energy spread of hydrogen atoms due to the movement of decaying neutrons and the angular divergence of the hydrogen "beam", the laser system including a resonator must be wide band, again imposing quasi-unrealistic conditions on the laser power required.\\
An alternate approach is a charge-exchange process of fast hydrogen atoms moving through a differentially pumped argon filled gas cell as $H + Ar \rightarrow H^- + Ar^+$. Negatively charged hydrogen atoms will be energy selected by an electric counter field (removing all low energy $H^-$-atoms) and momentum analyzed by means of a magnetic spectrometer.
\par
At a second stage of the experiment the hyperfine state of the hydrogen atom has to be determined. For this to happen the neutron must decay in a small magnetic field so the axis of polarization (magnetic quantum number $m_F$) is preserved. Outside the biological shield the hydrogen atoms pass through a spin filter (Lamb-shift spin-filter) also used for polarized hydrogen targets. By means of a B-field the $(F,m_F)$ states are Rabi split which leads to a crossing of the 2P state and one of the $2$S states. Using an RF field we can now depopulate specific 2S states which undergo a transition into a $2$S state crossing $2$P. In the presence of an electric field these two crossing states mix leading to a rapid quench into the $1$S ground state. After the spin filter, only the desired 2S state survives and must identified. 

\section{Conclusions}
80 years after the discovery of the neutron this particle species still offers a rich and very versatile laboratory for precision experiments addressing key questions in both, particle physics and cosmology. New sources for cold and ultra-cold neutrons are planned or currently going into operation offering new perspectives for "old" or new and innovative measurements by much improved intensity and/or beam quality. Besides the topics discussed in this short article, many more measurements are being performed with neutrons, a very prominent one being the determination of asymmetries or particle correlations in neutron decay (for an overview see 
\cite{abele2008}). But neutrons can also be used to study quantum mechanics through interference effects in crystals, to determine the quantization of electric charge by setting new limits on the electric charge of the neutron \cite{durstberger}, probing symmetry violating interactions or probing Lorentz invariance \cite{altarev2009}, search for mirror neutrons \cite{mirror_neutrons}  or baryon-antibaryon oscillations (baryon number violation) \cite{neutron_oscillations}. The physical significance of such measurements are mostly limited by the precision of the measurements themselves, rarely by uncertainties in theoretical corrections. However, most measurements hunt for deviations from Standard Model calculations. The interpretations of possibly observed effects thereby need complementary measurements, either with neutrons or in other systems (see e.g. EDM where atoms and electrons are also studied with equal importance). This field of science has seen a strong rise in attention with many new challenges being addressed. We can thus expect new and exciting highlights within the near future.


\begin{thebibliography}{99}
\bibitem{sakharov}A. D. Sakharov, JETP Lett.  {\bf 5} 24 (1967) 
\bibitem{Dubbers} D. Dubbers, M. Schmidt Rev. Mod. Phys. {\bf 83}, 11111171 (2011)

\bibitem{Nesvishevski} V. Nesvizhevsky  {\it et al.}, Phys. Rev. D  {\bf 67} 102002 (2003)
\bibitem{Abele03} H. Abele, S. Baessler, A. Westphal, Lecture Notes in Physics {\bf 631}, Springer (2003)  
\bibitem{Jenke11} T.Jenke, P. Geltenbort, H. Lemmel \& H. Abele, Nature Physics {\bf 7} 468-472 (2011)
\bibitem{Abele10}H. Abele, T. Jenke, H. Leeb, J. Schmiedmayer, Phys. Rev. D {\bf 81} 065019 (2010)
\bibitem{granit} V.V. Nesvizhevsky, Modern Phys. Lett. A {\bf 27(5)} 1230006 (2012)
\bibitem{Burgess2000}P. Callin and C.P. Burgess, "Deviations from Newtons Law in Supersymmetric Large Extra Dimensions", Nucl. Phys. B {\bf 752}, 60-79 (2006)
\bibitem{Kamyshkov2008} Y. Kamyshkov and J. Tithof, Phys. Rev. {\bf 78}, 114029 (6) (2008)

\bibitem{SM_EDM} X. He, B. H. McKellar, S. Paskvasa, Int. J. Mod. Phys. {\bf A4} 5011 (1989) 
\bibitem{ramsey57}J. H. Smith, E. M. Purcell and N. F. Ramsey, "Experimental limit to the electric dipole moment of the neutron", Phys. Rev. {\bf 108(1)} 120-122  (1957)
\bibitem{EDM_RAL}C. A. Baker, D. D. Doyle, P. Geltenbort, K. Green {\it et al.},  Phys. Rev. Lett. {\bf 97} 131801 (2006) 
\bibitem{fierlinger2012}S. Chesnevskaya {\it et al.},  
"A next generation measurement of the electric dipole moment of the neutron", Il Nuovo Cimento, (2012) {\it in press}
\bibitem{Frei2007}A. Frei {\it et al.}, Eur. Phys. J. {\bf A34} 119 (2007) 
\bibitem{EDM_PSI}I. Altarev {\it et al.},  Nucl. Instr. \& Meth. {\bf A611} 133-136 (2009) 
\bibitem{TRIUMF}see e.g. http://nuclear.uwinnipeg.ca/ucn/triumf
\bibitem{EDM_ILL}M.G.D. van den Grinten {\it et al.}, Nucl. Instr. \& Meth. {\bf A611} 129-132 (2009)  
\bibitem{EDM_SNS}T.M. Ito,  J.Phys.Conf.Ser. {\bf 69} :012037 (2007)

\bibitem{BBNS}Particle Data Group: C. Amsler {\it et al.}, Phys. Lett. B  {\bf 667}, 1 (2008), article by B.D. Fields and S. Sarkar\\
    K.A. Olive and E.D. Skillman Astrophys. J.  {\bf 617}, 29 (2004)
\bibitem{BBNS_2}G. Mathews, T. Kajino and T. Shima, Phys. Rev. D  {\bf 71}, 021302 (2005)
\bibitem{PDG06}Particle Data Group: W.-M. Yao {\it et al.}, J. of Phys. G  {\bf 33}, 1 (2006)


\bibitem{Arzumanov}S. Arzumanov {\it et al.}, Phys. Lett B  {\bf 483}, 15 (2000)
\bibitem{serebrov_1}A.P. Serebrov {\it et al.}, Phys. Lett. B  {\bf 605}, 72 (2005)  
\bibitem{Pichlmaier_10}A. Pichlmaier, V. Varlamov, K. Schreckenbach and P. Geltenbort
Phys. Lett. B, {\bf 693}:221-226 (2010)
\bibitem{PDG10}Particle Data Group: K. Nakamura {\it et al.},  J. Phys. G {\bf 37}, 075021 (2010) 
\bibitem{adelberger}E.G. Adelberger {\it et al.}, Rev. of Mod. Phys.  {\bf 70}, 1265 (1998)
\bibitem{declais}Y. Declais {\it et al.}, Phys. Lett. B  {\bf 338}, 383 (1994)
\bibitem{decay_formula}S. Weinberg, Phys. Rev.  {\bf 112}, 1375 (1958), M.L. Goldberger and S.B Treimann, Phys. Rev.  {\bf 111}, 354 (1958)
\bibitem{hardy08}I.S. Towner and J.C. Hardy, Phys. Rev. C  {\bf 77}, 025501 (2008)
\bibitem{Paul08} S. Paul, Nucl. Instr. \& Meth.  {\bf 611}, 157-166 (2009)
\bibitem{Picker} R. Picker {\it et al.},  J. Res. Natl. Inst. Stand. Technol. {\bf 110}, 357-360 (2005)
\bibitem{Leung} K.K. Leung and O. Zimmer, Nucl. Instr. \& Meth.  {\bf 611}, 181-185 (2009) 

\bibitem{goldhaber}M. Goldhaber, L. Grodzins and A. W. Sunyar, Phys. Rev. {\bf 109} 1015-1017 (1958)
\bibitem{twist}J.F. Bueno {\it et al.}, Phys. Rev. {\bf D84} 032005 (2011)
\bibitem{michel}L. Michel, Proc. Phys. Soc. {\bf A63} 514 (1950); C. Bouchiat
and L. Michel, Phys. Rev. {\bf 106} 170 (1957); T. Kinoshita
and A. Sirlin, Phys. Rev. {\bf 108} 844 (1957)
\bibitem{Nemenov} L.L. Nemenov, Sov. J. Nucl. Phys. {\bf 15}, 582 (1972)
\bibitem{Bahcall} J.N. Bahcall, Phys. Rev. {\bf124}, 495 (1961)
\bibitem{Song} X. Song, J. Phys. G: Nucl. Phys. {\bf 13}, 1023 (1987)
\bibitem{Schott} W. Schott {\it et al.}, EPJ A  {\bf 30}, 603 (2006)

\bibitem{abele2008}H. Abele, "The neutron. Its properties and basic interactions", Progr. Part. \& Nucl. Phys. {\bf  60-1} 1-81 (2008)
\bibitem{durstberger} see e.g. K. Durstberger-Rennhofer, T. Jenke, H. Abelear, Phys.Rev.  {\bf D84} :036004 (2011)
\bibitem{altarev2009}I. Altarev, Phys. Rev. Lett. {\bf 103}:081602 (2009)
\bibitem{mirror_neutrons}Z. Berezhiani, EPJ  {\bf C 64} 3, 421-431 (2009)
\bibitem{neutron_oscillations}R. N. Mohapatra,  J. Phys. G: Nucl. Part. Phys. {\bf 36} 104006 (2009)
\end{thebibliography}
\end{document}